 \documentclass[final,5p,times,twocolumn,authoryear]{elsarticle}

\usepackage{amssymb}
\usepackage{lipsum}
\usepackage{xcolor}
\usepackage{ulem}
\usepackage{neuralnetwork}
\usepackage{tikz}
\usetikzlibrary{shapes.geometric, arrows}
\usepackage{makecell}
\usepackage{newtxtext,newtxmath}
\usepackage{mathtools}

\usepackage{fmtcount}
\usepackage{array,multirow}
\usepackage{color}
\usepackage{multirow}
\makeatletter
\usepackage{xpatch}
\xpatchcmd{\linklayers}{\nn@lastnode}{\lastnode}{}{}
\xpatchcmd{\linklayers}{\nn@thisnode}{\thisnode}{}{}
\makeatother

\usepackage{hyperref}

\definecolor{dark-red}{rgb}{0.5,0.15,0.15}
\definecolor{dark-blue}{rgb}{0.15,0.15,0.5}
\definecolor{medium-blue}{rgb}{0,0,0.5}
\definecolor{medium-red}{rgb}{1,0,0}
\hypersetup{
    colorlinks={true}, linkcolor={medium-red},
    citecolor={dark-blue}, urlcolor={medium-blue}
}

\usepackage{xspace}
\newcommand{\teff}{{T_{\mathrm{eff}}}\xspace}

\newcommand{\logg}{\rm{log g}\xspace}
\newcommand{\feh}{\rm{[Fe/H]}\xspace}
\newcommand{\afe}{[\alpha/{\rm Fe}]}

\newcommand{\tgm}{\teff, \logg \text{ and } \feh}

\usepackage{xcolor, ulem}
\newcommand\hl{\bgroup\markoverwith{\textcolor{yellow}{\rule[-.5ex]{2pt}{2.5ex}}}\ULon} % highlighting command (more reliable than with soul)
\newcommand\hln{\bgroup\markoverwith{\textcolor{orange}{\rule[-.5ex]{2pt}{2.5ex}}}\ULon} % highlighting command (more reliable than with soul)
\journal{New Astronomy}

\begin{document}

\begin{frontmatter}



% \title{Title of paper}
% \title{Physical parameters of Stars in NGC 6397 using ANN based interpolation and Full Spectrum Fitting}

% \titlerunning{Parameter Estimation of Stars in NGC 6397 using ANN}


\title{Physical Parameters of Stars in NGC 6397 Using ANN-Based Interpolation and Full Spectrum Fitting}
\author[first]{Nitesh Kumar\texorpdfstring{\corref{cor1}}{}}
\ead{niteshchandra039@gmail.com}
\author[second]{Philippe Prugniel}
\author[third]{Harinder P. Singh}
\ead{hpsingh.du@gmail.com}
\affiliation[first]{organization={Department of Physics, University of Petroleum and Energy Studies (UPES)},
                     addressline={Bidholi}, 
                     city={Dehradun},
                     postcode={248007}, 
                     state={Uttarakhand},
                     country={India}}
\affiliation[second]{organization={Université de Lyon, Université Lyon 1},
                     addressline={69622 Villeurbanne}, 
                     city={Lyon},
                     postcode={69622}, 
                     state={Lyon},
                     country={France}}

\affiliation[third]{organization={Department of Physics and Astrophysics, University of Delhi},
                     addressline={North Campus}, 
                     city={Delhi},
                     postcode={110007}, 
                     state={Delhi},
                     country={India}}

% \fntext[label1]{Corresponding author: niteshchandra039@gmail.com}
\cortext[cor1]{Corresponding author: niteshchandra039@gmail.com}

\begin{abstract}
% Example abstract for the New Astronomy Journal. Here you provide a brief summary of the research and the results.

   {Stellar spectral interpolation is critical technique employed by fitting software to derive the physical parameters of stars. This approach is necessary because on-the-go generation of synthetic stellar spectra is not possible due to the complex and high cost of computation. }
   {The goal of this study is to develop a spectral interpolator for a synthetic spectral library using artificial neural networks (ANNs). The study aims to test the accuracy of the trained interpolator through self-inversion and, subsequently, to utilize the interpolator to derive the physical parameters of stars in the globular cluster NGC 6397 using spectra obtained from the Multi Unit Spectroscopic Explorer (MUSE) on the Very Large Telescope (VLT).}
   {In this study, ANNs were trained to function as spectral interpolators. The ULySS full-spectrum fitting package, integrated with the trained interpolators, was then used to extract the physical parameters of 1587 spectra of 1063 stars in NGC 6397.}
   {The trained ANN interpolator achieved precise determination of stellar parameters with a mean difference of 31 K for $\teff$ and 0.01 dex for $\feh$ compared to previous studies.}
   {This study demonstrates the efficacy of ANN-based spectral interpolation in stellar parameter determination, offering faster and more accurate analysis.}

\end{abstract}




\end{frontmatter}

%\tableofcontents



\section{Introduction}\label{introduction}

Stellar spectral libraries are fundamental tools in astrophysics, providing reference spectra for various applications. They are essential for stellar classification \citep{Buser_1992}, the determination of stellar parameters \citep{Lejeune_1997A&AS..125..229L}, and the calibration of spectroscopic surveys \citep{Lejeune_1998A&AS..130...65L}. These libraries also play a crucial role in stellar population synthesis and galaxy evolution studies \citep{Westera_2002A&A...381..524W}, as well as in refining theoretical stellar atmosphere models \citep{Conroy_2013}. Stellar spectral libraries comprise collections of stellar spectra along with the corresponding atmospheric parameters and stellar metallicities. Empirical spectral libraries such as Indo-US Library of Coudé Feed Spectra \citep{valdes_indo-us_2004}, MILES \citep{sanchez-blazquez_medium-resolution_2006}, ELODIE \citep{prugniel_new_2007} and the X-shooter library \citep{verro_x-shooter_2022} feature observed spectra that span a broad range of parameters across the Hertzsprung-Russell (H-R) diagram. These libraries have been extensively employed for stellar spectral classification \citep{gulati_stellar_1994, bailer-jones_automated_1998, singh_stellar_1998, navarro_automatic_2012, liu_spectral_2015, sharma_application_2020} and for determining stellar parameters e.g., \cite{prugniel_atmospheric_2011, wu_coude-feed_2011, sharma_new_2016}. They have also been utilized in chemical and evolutionary studies of stellar populations \citep{leitherer_starburst99_1999, crowl_stellar_2008}. Despite capturing real physical features, these libraries often have limitations in terms of spectral wavelength coverage and resolution. Additionally, they encompass a restricted range of parameters, and the precise chemical compositions of the observed stars are typically not known, which can lead to systematic errors in abundance determinations. 

Synthetic spectra are generated by modelling stellar atmospheres (such as ATLAS \citep{kurucz_synthe_1993}, PHOENIX \citep{hauschildt_numerical_1999}, MARCS \citep{gustafsson_grid_2008}), incorporating up-to-date line list and accounting for line broadening parameters. These spectra are computationally synthesized using advanced software, enabling the creation of a collection that spans a broad range of atmospheric parameters. The resulting spectra can be of specific spectral resolutions and wavelength coverage. However, producing synthetic spectra is computationally intensive and time-consuming due to the complexity of the models and the detailed calculations required.

Despite these challenges, several synthetic spectral libraries such as \cite{munari_extensive_2005}, \cite{husser_new_2013} and \cite{coelho_new_2014} have been computed. These libraries offer a wide variety of spectra that cover diverse regions of parameter space, including different effective temperatures, surface gravities, and chemical compositions. Synthetic spectral libraries are advantageous because they provide the flexibility to generate spectra under conditions that may not be easily observable, such as extreme metallicities or very high surface gravities.

In full spectral fitting methods, stellar atmospheric parameters are determined by comparing the observed spectrum to interpolated spectra from a grid of empirical or synthetic spectral libraries using the $\chi^2$ minimization technique. Several interpolation techniques have been used previously, including linear, quadratic and cubic Bezi\'er splines \citep{auer_formal_2003} for FERRE \citep{allende-prieto_ferre_2015} and polynomial-based interpolation schemes \citep{prugniel_atmospheric_2011, sharma_new_2016} for ULySS \citep{koleva_ulyss_2009}.

Spectral interpolation has traditionally been performed using a polynomial function of $\teff, \logg, \feh$, and $\lambda$ \citep[see Equation 3 in][]{prugniel_atmospheric_2011}. In this approach, polynomial terms are iteratively selected, with the final combination determined by minimizing the residuals between the original and interpolated spectra in the spectral library. The number of terms required depends on factors such as spectral resolution, wavelength coverage, and the range of atmospheric parameters in the stellar spectral library. However, developing such polynomial-based interpolators is a labor-intensive process, often requiring fine-tuning or adding terms to handle specific star types or parameter space edges. An improved version was introduced by \cite{sharma_new_2016} to address these challenges, particularly for cool stars, which are crucial for galactic studies \citep{chabrier_galactic_2003, west_using_2006} and exoplanetary research \citep{bonfils_metallicity_2005, neves_metallicity_2013}.

Machine learning (ML) methods provide a powerful alternative, eliminating the need to manually develop polynomials for mapping atmospheric parameters ($\teff, \logg, \feh$) to flux values. Instead, ML models automatically capture and learn these mappings directly from the dataset. ML-based approaches are more flexible, generalizable, and effective across various spectral libraries. \cite{sharma_stellar_2020} showed that ML models provides better accuracy compared to traditional polynomial-based interpolation and Gaussian radial basis function (RBF) methods, making them a promising tool for spectral interpolation.

ML methodologies have seen extensive application in the realm of astronomical data analysis, encompassing both spectroscopic and photometric datasets in recent times. ML techniques can generally be categorized into four main types: supervised learning, unsupervised learning, semi-supervised learning, and reinforcement learning. Supervised ML algorithms require a set of input features along with corresponding labels for training purposes. These algorithms ``learn" by optimizing the free parameters or ``weights" based on the provided training data. Supervised learning is capable of performing classification tasks, where the features are associated with discrete label values, and regression tasks, where the features are correlated with continuous value(s). Such algorithms have been effectively utilized in the analysis of stellar spectra \citep{bailer-jones_automated_1998, solorio_active_2005, sharma_stellar_2020} in time-domain astronomy \citep{miller_machine-learning_2015, sedaghat_effective_2018, sanchez_machine_2019}, for the interpolation of light curves of RRab stars \citep{kumar_predicting_2023}, in deriving the physical parameters of RRab stars \citep{kumar_multiwavelength_2024}, and in determining galaxy morphology and other parameters \citep{ball_galaxy_2004, abraham_detection_2018}.

Supervised ML algorithms have been used in the past for stellar spectral interpolation. \cite{sharma_stellar_2020} employed machine learning (ML) algorithms, specifically artificial neural networks (ANN) and random forests (RF), to demonstrate the feasibility of spectral interpolation. They trained the ML models to map the relationship between three atmospheric parameters ($\teff$, $\logg$, and $\feh$) and the corresponding spectra from the MILES and Indo-US spectral libraries. A similar approach was given by \cite{ness_cannon_2015}, where they presented a probabilistic generative model named \textit{The Cannon}, where they learned the mapping of atmospheric parameters ($\tgm$) with the continuum normalized stellar spectra. A probability density function with mean and variance at each wavelength bin for each spectrum in the reference library is generated for the model.   
This learned mapping is then used to determine the physical parameters of spectra for which the parameters are not known. They employed high-resolution (R $\sim$ 22500) APOGEE \citep{majewski_apache_2017} spectra in the IR domain (15200 - 16900 \AA{}) as training and test samples. The $\teff$ coverage of these stars was 3500 - 5500 K. \cite{dafonte_estimation_2016} used a generative artificial neural network for generating the stellar spectra using four input parameters $\teff, \logg, \feh$ and $\afe$ in the caII triplet region (847-871 nm). For training, they utilized the medium resolution (R $\sim$ 11200) synthetic spectra with $\teff$ in the range 4000 - 11 500 K and $\logg$ in the range 2.0 - 5.0 dex. The spectra from \textit{Gaia} Radial Velocity Spectrograph (RVS) instrument were used to estimate the stellar parameters. \cite{cheng_new_2018} introduced the approach to interpolate stellar spectra using a radial basis function (RBF) network. They used the MILES spectral library for training the network in the optical region. \textit{The Payne}, trained by \cite{ting_payne_2019}, generates the stellar spectra in the APOGEE wavelength region using 25 stellar labels, including stellar physical parameters and individual element abundances. The network was trained with 2000 synthetic spectra of giants and dwarfs in the $\teff$ range of 3000-8000 K.        

In this study, we developed a spectral interpolator utilizing artificial neural networks (ANNs), leveraging their capability as function approximators for continuous and differentiable functions \citep{cybenko_approximation_1989, hornik_multilayer_1989}. The medium-resolution G\"{o}ttingen spectral library \citep[GSL;][]{husser_new_2013} was employed to train the ANNs, with parameters $\teff$, $\logg$, and $\feh$ parameters serving as inputs, and the corresponding spectra serving as outputs. We generated the necessary `interpolator' files required for the ULySS software and implemented new routines to integrate the ANN-based spectral interpolators into the ULySS framework.

Globular clusters (GCs) host a large number of stars which are useful for our understanding of stellar populations, stellar evolution, and the chemical compositions of the stars within them. With the introduction of wide-field integral field spectrographs such as MUSE \citep{bacon_muse_2010, bacon_muse_2014} mounted at the Very Large Telescope (VLT), it has been possible to get the medium resolution spectra of sources (above a certain brightness limit) in the cluster with reasonable pointings. It was used by \cite{husser_muse_2016} (hereafter H16) to observe the NGC 6397 (see details in section \ref{subsec:ngc6397}). We used the 1587 spectra of 1063 stars above the turn-off (TO) point of the main sequence to derive the stellar parameters using the ANN spectral interpolator and then compared the results with the existing literature. 

The structure of this paper is as follows: In Section \ref{sec:data}, we discuss the spectral data utilized for training the network, including the MUSE spectra of stars in NGC 6397, for which atmospheric parameters were derived using the ANN spectral interpolator. Section \ref{sec:methods} described the details of the artificial neural network model used. In Section \ref{sec:ANN_training}, we provide the training process of the ANN models and describe the self-inversion of the GSL spectra to evaluate the accuracy of the derived parameters using the ANN spectral interpolator within the ULySS. In Section \ref{sec:ngc6397_physical_parameters}, we determine the physical parameters of stars in NGC 6397 from their MUSE spectra, employing full-spectral fitting methods with the ANN spectral interpolator trained on the GSL grid. In Section \ref{sec:discussion}, we discuss the metallicity of NGC 6397, and finally, in Section \ref{sec:conclusion}, we summarize this study.

%-------------------------------------------------------------------
\section{Data}\label{sec:data}
\subsection{Training Data}\label{subsec:training_data}
The G\"{o}ttingen spectral library (GSL), as detailed in \citep{husser_new_2013}, constitutes a compilation of synthetic stellar spectra generated using the PHOENIX synthesis code \citep{hauschildt_nextgen_1999}. This repository encompasses both high and medium-resolution versions. In our study, we specifically employed the medium-resolution dataset (R = \( \lambda/\Delta\lambda = 10,000 \)), featuring an alpha-element abundance of \( \afe = +0.4 \). This choice is particularly suitable for analyzing the stellar spectra of NGC\,6397, which has an average alpha-element abundance of $ \afe = +0.36 $, as determined by \citet{carretta_na-o_2009, carretta_properties_2010} using UVES spectra. The wavelength coverage is from 3000\AA{} to 25000\AA{} with bins evenly distributed on a logarithmic scale. The total input parameter space is shown in Table \ref{tab:gsl_table}. We did not use the extension to higher temperatures released after the initial publication. In total, the dataset counts 2908 spectra out of the 3198 combinations of the three parameters inferred from this table because the coverage is incomplete at the edge of the parameter space and because the computation of some spectra did not converge.

% \begin{figure*}
%     \centering
%     \includegraphics{figs/input_grid_params.pdf}
%     \caption{The distribution of physical parameters ($\teff$, $\logg$ and $\feh$) for the medium resolution (R $\sim$ 10 000) G\"ottingent spectral library for $\afe$ = +0.40.}
%     \label{fig:input-params}
% \end{figure*}

To adapt the data to the present need, we preprocessed the spectra by applying the following procedures:

\begin{enumerate}
    \item We reduced the resolution of spectra from $\sim$ 0.6 \AA{} (at 6000 \AA{}) to 2 \AA{} by convolving the original spectrum with a Gaussian of FWHM = $\sqrt{2^2 - 0.6^2}$,
    \item We re-binned the spectra at wavelength bins of 1.25 \AA{} by first integrating the spectra and then differentiating at the boundaries of wavelength bins,
    \item We trimmed the wavelength range to 4762.25-7384.75 \AA{} to mitigate the telluric lines, 
    \item Finally, we normalised the spectra by dividing each spectrum by its median flux value. 
\end{enumerate}

The above steps result in a training set where each spectrum consists of 2099 flux values, covering the wavelength range from 4762.25 to 7384.75 \AA{} with a bin size of 1.25 \AA{}. This pre-processing step is essential for matching the spectral resolution of standard medium-resolution spectrographs in the optical wavelength region. By simplifying the training set in this manner, the training process becomes more efficient and rapid, facilitating improved model performance and computational speed. 

\begin{table}
	\centering
	\caption{Parameter space of medium resolution GSL.}
	\label{tab:gsl_table}
	\begin{tabular}{lccr} % four columns, alignment for each
		\hline \hline
		Parameter    & Range        &  Step-size\\
		\hline
		$\teff$[K]  & 3500 - 7000  & 100 \\
		              & 7000 - 8000  & 200 \\
		$\logg$     &  0.0 - 6.0   & 0.5 \\
		
		$\feh$      & -3.0 - -2.0  & 1.0 \\
		              & -2.0 - +0.0  & 0.5 \\
		$\afe$      &  +0.4        & -- \\
		\hline
	\end{tabular}
\end{table}

\subsection{NGC 6397}\label{subsec:ngc6397}

NGC 6397 is a Globular Cluster (GC) located at a distance of 2.48 kpc \citep{baumgardt_accurate_2021} from the Sun, making it the second nearest GC after Messier 4. It is an old (12.6 Gyr old; \citealt{correnti_age_2018}), metal-poor ([Fe/H] = -2.02; \citealt{harris_catalog_1996, harris_new_2010}) GC. The spectroscopic observations of the NGC 6397 were conducted during the commissioning phase of Multi Unit Spectroscopic Explorer (MUSE\footnote{\url{https://www.eso.org/sci/facilities/develop/instruments/muse.html}}; \citealt{bacon_muse_2014}) from 26 July to 3 August 2014 by H16. MUSE is an integral field spectrograph installed at UT4 (Yepun) of the ESO Very Large Telescope (VLT) in Chile. It provides a wide field of view of $1 \times 1$ arcmin$^2$ with a spatial sampling of $0.2 \times 0.2$ arcsec per pixel in the Wide Field Mode (WFM). It covers a wavelength range from 480 to 930\,nm, delivering a mean spectral resolution of $\sim3000$. MUSE consists of 24 identical integral-field units, enabling the simultaneous acquisition of over 90,000 spectra in a single exposure \cite{bacon_muse_2010}.

A $5\times5$ mosaic was planned to cover the central region of the NGC 6397 cluster, extending to approximately 3.5 arc-minutes from the centre. However, due to interruptions for instrument testing, two fields in the peripheral regions could not be observed. The total integration time for the 127 exposures amounted to 95 minutes. The seeing conditions during these observations consistently remained below approximately 1 arc-second.

% The data reduction were done using the official MUSE pipeline \citep{weilbacher_data_2020}. The de-blending technique used in the extraction of individual spectra was developed by \cite{kamann_resolving_2013}.  A total 18 932 spectra of 10 521 sources is extracted from the raw data. The 14 271 spectra with signal-to-noise ratio (SNR) greater than 5 are provided on public archive. The line spread function (LSF) varies with the wavelength from full width at half maximum of about 2.82 \AA{} at $\lambda = 4750$ \AA{} to about $2.54$ \AA{} at $\lambda = 7000$ \AA{}. 

Data reduction was performed using the official MUSE pipeline \citep{weilbacher_data_2020}. The de-blending technique employed for extracting individual spectra was developed by \cite{kamann_resolving_2013}. The stellar spectra of the individual stars and the description of the observations are provided in H16. From the raw data, a total of 18932 spectra corresponding to 10 521 sources were extracted. Among these, 14271 spectra with SNR greater than 5 are available in the public archive. The line spread function (LSF) varies with wavelength, having a full width at half maximum (FWHM) of approximately 2.82 \AA{} at $\lambda = 4750$ \AA{} and about 2.54 \AA{} at $\lambda = 7000$ \AA{}.

H16 derived the atmospheric parameters of 4132 stars from 5881 spectra (as there are overlapping individual mosaics and the same star is observed multiple times) with S/N $>$ 20. They adopted a three-step process: (i) $\teff$ and $\logg$ were determined by fitting Hubble Space Telescope (HST) photometry to an isochrone, (ii) the radial velocity ($v_{\rm rad}$) was obtained by cross-correlating interpolated models generated using the parameters derived in the first step, (iii) final optimization was conducted to refine the values of $\teff$, $\feh$, $v_{\rm rad}$, line broadening, and the telluric absorption spectrum. The GSL spectra, interpolated using the cubic spline method, served as the reference for their analysis. To prevent possible degeneracy between $\logg$ and broadening, they did not optimize $\logg$, instead adopted the photometric $\logg$ values in their results.

\cite{jain_ngc_2020} conducted a reanalysis of the spectra from H16 and derived the physical parameters of the stars using the ELODIE and MILES spectral libraries. They further refined their sample by including only those stars that were identified as cluster members based on radial velocity measurements and also omitted hot stars with effective temperatures ($\teff$) exceeding 7000 K. Additionally, they limited their analysis to stars with surface gravity ($\logg < $)  4.2. This selection resulted in a final dataset comprising 1587 spectra from 1063 stars.

The same set of stars (except three spectra of two stars which have $\logg <$ 1) were again analysed by \cite{baratella_prospects_2022} (hereafter B22) using a different software named FERRE with two distinct synthetic spectral libraries AP18 \citep{allende_prieto_collection_2018} and GSL using various interpolation schemes with varying $\afe$. There are two fundamental differences between the FERRE and ULySS: (i) The FERRE requires a homogeneous grid of synthetic model spectra and does not work if there is a single missing node in the parameter space, (ii) Unlike ULySS, the FERRE requires the model and observed spectrum to have the same line spread function. In ULySS, the broadening function and systematic velocity are fitted at the same time as the physical parameters with a multiplicative polynomial that takes care of the improper flux calibrations and extinctions. 

\section{Methodology}\label{sec:methods}

\begin{figure}
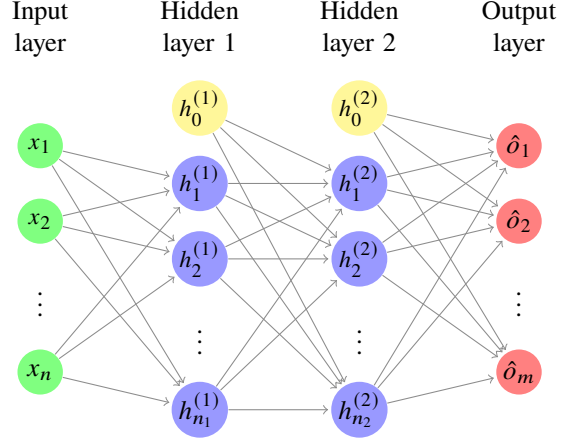

	\begin{center}
		\begin{neuralnetwork}[height=5, style={}, title={}, titlestyle={}, layertitleheight=1.05cm, layerspacing=2.1cm]

			% use \ifnum to get different labels
			\newcommand{\x}[2]{\ifnum #2=4 $x_n$ \else \small $x_#2$ \fi}
			
			\newcommand{\hfirst}[2]{\ifnum #2=4 $h^{ (1)}_{n_1}$ \else \small $h^{ (1)}_#2$ \fi}
			\newcommand{\hsecond}[2]{\ifnum #2=4 $h^{ (2)}_{n_2}$ \else \small $h^{ (2)}_#2$ \fi}

			\newcommand{\nodetexty}[2]{\ifnum #2=4 $\hat{o}_m$ \else $\hat{o}_#2$ \fi}
			
			% use exclude to turn off drawing of specific layers
			\inputlayer[count=4, bias=false, exclude={3}, title=Input\\layer\\, text=\x]
		
			\hiddenlayer[count=4, bias=true, exclude={3}, title=Hidden\\layer 1, text=\hfirst] \linklayers[not to={3}, not from={3}]

			\hiddenlayer[count=4, bias=true,exclude={3}, title=Hidden\\layer 2, text=\hsecond] \linklayers[not to={3}, not from={3}]
			
			\outputlayer[count=4, exclude={3}, title=Output\\layer, text=\nodetexty] \linklayers[not to={3},not from={3}]
			
			% draw dots
			\path (L0-2) -- node{$\vdots$} (L0-4);
			\path (L1-2) -- node{$\vdots$} (L1-4);
			\path (L2-2) -- node{$\vdots$} (L2-4);
			\path (L3-2) -- node{$\vdots$} (L3-4);
		\end{neuralnetwork}
	% \caption{A schematic representation of an artificial neural network with $n$ input nodes, two hidden layers containing $n_1$ and $n_2$ neurons respectively, and an output layer comprising $m$ neurons. The bias terms for each respective layer are denoted as $h_0$. For the sake of clarity, summation and non-linearity nodes are not depicted.}
    \caption{A schematic representation of an artificial neural network used for spectral prediction. The input layer consists of stellar parameters from the GSL grid, the hidden layers process this information, and the output layer generates the predicted spectral fluxes. The bias terms for each respective layer are denoted as $h_0$. Summation and non-linearity nodes are omitted for clarity.}
	\label{fig:MLP}
	\end{center}
\end{figure}

To construct an efficient spectral interpolator, we employ artificial neural networks (ANNs) to model the relationship between stellar atmospheric parameters and spectral fluxes. The ANN learns to map the input parameters ($\teff$, $\logg$, $\feh$) to the corresponding spectra, enabling smooth and accurate interpolation across the GSL grid.
We utilize a basic neural network model: a feed-forward, fully-connected neural network. Each fundamental unit, known as a neuron or a perceptron, performs a mathematical operation by computing the weighted sum of all its connected neurons from the previous layer, then applies a non-linear activation function ($\sigma$) before passing the result to the next layer (see Fig. \ref{fig:MLP}). The value of $i^{th}$ neuron in $k^{th}$ layer is determined as follows:

\begin{equation}\label{eq:single_neuron}
    h_i^{ (k)} = \sigma^{(k)} \left (\sum_{j=0}^{n_{k}-1} w_{ij}^{(k-1)} h_{j}^{ (k-1)} \right),\: \text{for} \: 1 \leq i \leq n_{k},
\end{equation}    
with,
\begin{equation}
h_0^{ (k)} \equiv 1 ,  
\end{equation}
here $\sigma^{(k)}$ is the activation function for the $k^{th}$ layer, typically a non-linear function such as rectified linear unit (ReLU), sigmoid, or hyperbolic tangent (tanh). The weights connecting the $j^{th}$ neuron to the $i^{th}$ neuron in the $k^{th}$ layer are represented by $w_{ij}^{(k)}$, and these weights are optimized to minimize (or maximize) a specified objective function during the training process.

Weight updates occur iteratively through the backpropagation algorithm, which was introduced by \cite{rumelhart_learning_1986}.
This algorithm propagates errors from the output layer back through the network, improving performance with each iteration. Backpropagation underpins modern stochastic gradient descent (SGD) algorithms, which optimize network weights to achieve the desired training objective \citep[for a comprehensive review of modern SGD algorithms, see][]{ruder_overview_2016}.

Figure \ref{fig:MLP} illustrates a network schematic with two hidden layers. For the input layer, we set $k \equiv 0$ and $\bf{h}^{(0)} \equiv \textbf{x}$ and for the output layer, $k \equiv L$ and ${\textbf{h}}^{ (\text{L})} \equiv \hat{\bf{o}}$ (where $\hat{o}$ is the network output).

To evaluate the model, we use the mean square error (MSE) as the objective function. For the $i^{th}$ model, the MSE is calculated as:

\begin{equation}\label{eq:avg_mse}
    \mathbb{E}_{i} \equiv (\text{MSE})_{i} = \frac{1}{N_s} \sum_{j=1}^{N_s} (o_{ij} - \hat{o}_{ij})^2,
\end{equation} 
where $\hat{o}_{ij}$ and $o_{ij}$ represent the predicted and actual absolute magnitude values for the $i^{th}$ model at the $j^{th}$ phase, respectively, and $N_s$ is the total number of magnitude bins per model. The average MSE across all models in the dataset is computed as:
\begin{equation}
\textbf{$\mathbb{E}$} \equiv \text{Avg. MSE} = \frac{1}{N} \sum_{i=1}^{N} (\text{MSE})_{i} = \frac{1}{N} \sum_{i=1}^{N} \left ( \frac{1}{N_s} \sum_{j=1}^{N_s} (o_{ij} - \hat{o}_{ij})^2 \right). 
\end{equation}
where $N$ is the total number of models in the training dataset.

During each training iteration, for simple gradient descent, weights are updated as follows:

\begin{equation}\label{eq:weight_update}
    w_{ij}^{(k)} = w_{ij}^{(k)} - \eta \times \frac{\partial \textbf{$\mathbb{E}$}}  {\partial w_{ij}^{ (k)}}
\end{equation}
where $\eta$ is the learning rate that controls the step size of gradient descent. However, we employ a variant of gradient descent known as the adaptive moment estimation (Adam) optimizer \citep{kingma_adam_2014}, which adaptively adjusts the learning rate at each iteration to efficiently minimize the objective function.

%-------------------------------------------------------------------

\section{ANN Training and Self Inversion}\label{sec:ANN_training}

\subsection{Training of the ANN Interpolator}
We trained several artificial neural networks (ANNs) with varying architectures. Each architecture features an input layer with 3 neurons representing the atmospheric parameters, and an output layer comprising 2099 neurons, each corresponding to a flux value in a specific wavelength bin within the range 4762.25\AA{} to 7384.75\AA{}. Hidden layers utilize a logistic (sigmoid) activation function, while the output layer does not. The output layer utilizes the linear activation function. The three input parameters, effective temperature ($\teff$), surface gravity ($\logg$), and metallicity ($\feh$), exhibit different scales and hence require pre-processing to ensure the stability of the neural network. To standardize these parameters, we computed the mean and standard deviation for each one. Each parameter value was then adjusted by subtracting the mean and dividing by the standard deviation. This process results in a dataset where each parameter has a mean of zero and a standard deviation of one, ensuring uniform scaling across all three dimensions in parameter space. This technique is commonly referred to as `Standard Scaling' in the field of machine learning. 

\begin{table*}
    \centering
    \caption{The different network architectures are shown along with the training time and minimum mean squared error achieved by the model.}
    \label{tab:train_table}
    \begin{tabular}{c  r  r  r  r  r  r }
\hline \hline
Trial No.     &     \makecell[t]{Network \\ Architecture} & \makecell[t]{Total \\ params} &  \makecell[t]{Params \\ per bin}    &  Epochs &  \makecell[t]{Time \\ (s)} &     Min MSE  \\

\hline 
0         &        [3, 32, 128, 512, 2099] &       1147187 &             547 &     314  &       316 &  $4.95 \times 10^{-5}$\\
1         &        [3, 32, 128, 512, 2099] &       1147187 &             547 &     275  &       286 &  $5.77 \times 10^{-5}$\\
2         &  [3, 32, 128, 512, 1024, 2099] &       2747187 &            1309 &     273  &       712 &  $4.75 \times 10^{-5}$\\
3         &       [3, 64, 256, 1024, 2099] &       2431539 &            1158 &     270  &       605 &  $4.28 \times 10^{-5}$\\
4         &       [3, 64, 256, 1024, 2099] &       2431539 &            1158 &     389  &       966 &  $3.35 \times 10^{-5}$ \\
5         &       [3, 64, 256, 1024, 2099] &       2431539 &            1158 &  $10^3$  &      2444 &  $2.11 \times 10^{-5}$\\
6         &        [3, 32, 128, 512, 2099] &       1147187 &             547 &  $10^5$  &     99954 &  $6.20 \times 10^{-7}$\\
7         &              [3, 32, 64, 2099] &        138675 &              66 &  $10^5$  &     23839 &  $8.12 \times 10^{-6}$\\
8         &              [3, 32, 64, 2099] &        138675 &              66 &  $10^6$  &     232523&  $4.77 \times 10^{-6}$ \\
\hline
% \bottomrule
\end{tabular}
\end{table*}

We employed the average mean squared error (MSE) as the objective function (see Equation \ref{eq:avg_mse}), described in Section \ref{sec:methods}, and updated the network parameters iteratively to minimize MSE using the adaptive moment estimation (Adam) optimization algorithm \citep{kingma_adam_2014}, with an initial learning rate of 0.001. The network parameters were initialized according to the Glorot normal distribution \citep{glorot_understanding_2010}. The batch size was set to the default value of 32.

We conducted nine different training trials, each with a distinct network architecture and varying training iterations. For the first five trials, we used early stopping criteria to halt the training when the loss (MSE) value stopped decreasing for 50 consecutive epochs. For the rest of the networks, we relaxed this criterion and trained the network for longer epochs. For each architecture, we calculated the total number of network parameters, which serves as an indicator of the model's complexity. We also computed the `parameters per bin' by dividing the total network parameters by the number of wavelength bins (the number of neurons in the output layer). This metric reflects the compression factor for the given grid. 

Table \ref{tab:train_table} details the network architecture (which has the number of atmospheric parameters as first value, the number of spectral pixels at the last and the number of neurons in the hidden layer in between), number of epochs, training duration (computed on a laptop equipped with an NVIDIA GeForce MX150 GPU with 4GB memory and 384 CUDA cores), and the minimum MSE achieved during training. 

\begin{figure}
    \centering
    \includegraphics[scale=0.80]{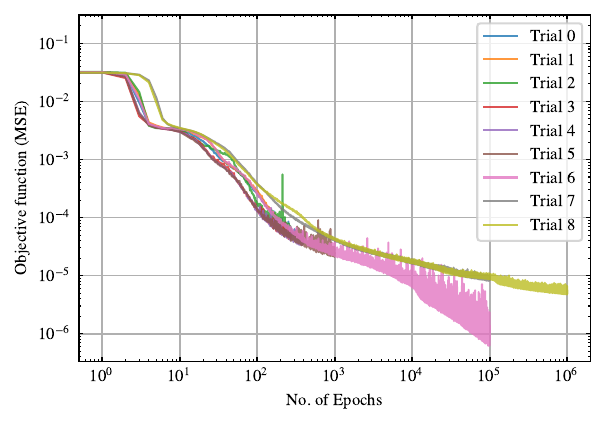}
    \caption{This figure shows the path of optimisation of the objective function along with the epochs of training.}
    \label{fig:loss_vs_epoch}
\end{figure}

\subsection{Inversion: Parameter determination using ANN Interpolator}
Parameter determination involves a two-step process. Initially, a grid of synthetic (or empirical) stellar spectra is interpolated within the atmospheric parameter space (training the interpolator). In the second step, the observed spectrum (the spectrum which has to be parameterised) is compared with the spectrum generated through the interpolated model. A minimisation scheme (typically $\chi^2$ minimisation) is performed to retrieve the best-fit parameters. There exists a number of softwares that can perform the same task. Some of the examples include FERRE, Spexxy\footnote{\url{https://github.com/thusser/spexxy}} and ULySS. We utilized the University of Lyon Spectroscopic analysis Software \citep[ULySS\footnote{\url{http://ulyss.univ-lyon1.fr/}},][]{koleva_ulyss_2009} for the inversion process. ULySS is an open-source software package, written in GDL/IDL, designed for analyzing spectroscopic astronomical data. It fits a spectrum using a linear combination of non-linear components convolved with a line-of-sight velocity distribution (LOSVD) and multiplied by a polynomial continuum \cite{koleva_ulyss_2009}. This software is widely used for studying stellar populations in galaxies and star clusters, as well as determining the atmospheric parameters of stars.

ULySS can generate a polynomial-based interpolator from a given set of empirical spectral libraries such as ELODIE \citep{prugniel_database_2001} and MILES \citep{sanchez-blazquez_medium-resolution_2006}. The parameters of the stars of the MILES spectral library are determined using ELODIE as reference in \cite{prugniel_atmospheric_2011}. The ANN interpolator trained in this work has been incorporated with the ULySS, and the useful files required to use the generated ANN interpolator are provided at GitHub repository \href{https://github.com/niteshchandra039/ANN_spectral_Interpolator}{https://github.com/niteshchandra039/ANN\_spectral\_Interpolator}. 

For each trained interpolator, we retrieve the atmospheric parameters from the grid spectra using ULySS. The precision of the interpolator is quantified by the average residual variance between the actual parameters and the inferred parameters. 

The absolute precision for each atmospheric parameter at a given node is calculated by taking the difference between the actual values and those retrieved using ULySS. The overall precision for each parameter across the grid is determined by calculating the square root of the average squared distance between the inverted and actual parameters in parameter space (see equation \ref{eq:precision}). This absolute precision is reported in Table \ref{tab:inversion_table} for the GSL grid. To further quantify precision, we express it as a percentage of the mesh size recoverable by the interpolator, calculated by dividing the absolute precision by the average step size of the parameter in the grid.

\begin{equation}\label{eq:precision}
    P_X = \sqrt{\frac{\sum_i^N(X_{\rm derived, i} - X_{\rm actual, i})^2}{N}}
\end{equation}

where $X_{\rm derived, i} \in [\tgm]$ is derived using ANN interpolator with ULySS and $X_{\rm actual, i}$ are the original parameters from the training grid and $N$ is the number of training samples.

\begin{table*}
    \centering
    \caption{The absolute and percentage precision of the GSL grid for all architectures are presented in this table.}

\resizebox{\textwidth}{!}{
\begin{tabular}{l c  c  r  c  c  c  c  c  c  c  c  c  c }     % 7 columns 
\hline\hline
Trial no.              & Network                         & Epochs               & Min MSE               & Time                 & \multicolumn{3}{c}{Absolute Precision}                             & \multicolumn{3}{c}{Percentage Precision}                            \\
                       & Architecture                    &                      &                       & (s)                  & $P_{\teff}$ (K)                  &     $P_{\logg}$ (dex)                 & $P_{\feh}$ (dex)                     & $P^{'}_{\teff}$ (\%)                 & $P^{'}_{\logg}$ (\%)                 & $P^{'}_{\feh}$ (\%)                  \\
                       &                                 &                      &                       &                      &                      &                      &                      &                      &                      &                       \\ 
\hline

0         &        [3, 32, 128, 512, 2099] &       314 &  4.95$ \times 10^{-5} $ &       316 &    49 &  0.28 &      0.24 &  41 & 52 & 38\\
1         &        [3, 32, 128, 512, 2099] &       275 &  5.77$ \times 10^{-5} $ &       286 &    59 &  0.37 &      0.23  & 49 & 70 & 36 \\
2         &  [3, 32, 128, 512, 1024, 2099] &       273 &  4.75 $ \times 10^{-5} $ &       712 &    58 &  0.25 &     0.15  & 49 & 46 & 23 \\
3         &       [3, 64, 256, 1024, 2099] &       270 &  4.28 $ \times 10^{-5} $ &       605 &    49 &  0.23 &     0.16  & 41 & 42 & 25 \\
4         &       [3, 64, 256, 1024, 2099] &       389 &  3.35$ \times 10^{-5} $ &       966 &    32 &  0.18  &     0.11  & 27 & 34 & 16 \\
5         &       [3, 64, 256, 1024, 2099] &     $10^3$ &  2.11$ \times 10^{-5} $ &      2444 &    25 &  0.16  &     0.10 & 21 & 30 & 16 \\
6         &        [3, 32, 128, 512, 2099] &    $10^5$ &  6.20$ \times 10^{-7} $ &     99954 &     6  &  0.03  &     0.02 &  5 & 6 &  3 \\
7         &              [3, 32, 64, 2099] &    $10^5$ &  8.12$ \times 10^{-6} $ &     23839 &    20 &  0.09  &     0.06  & 16 & 16 & 10 \\
8         &              [3, 32, 64, 2099] &    $10^6$ &  4.77$ \times 10^{-6} $ &    232523 &    15  &  0.07 &     0.04  & 13 & 12 & 6 \\
\hline                  
\end{tabular}
}
    \label{tab:inversion_table}
\end{table*}

\begin{figure}
    \centering
    \includegraphics[scale=0.8]{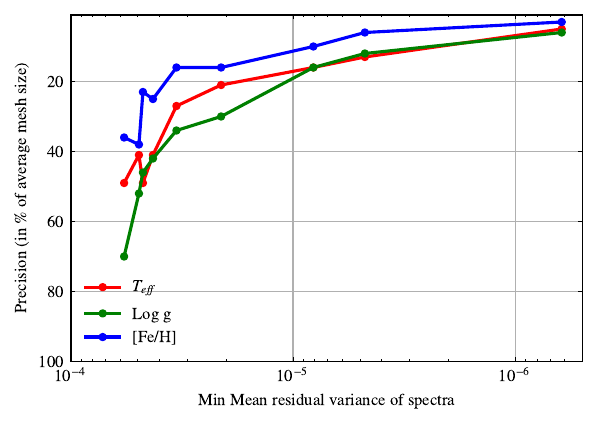}
    \caption{The precision in \% of the average mesh size of the grid is plotted against the minimum mean squared error the network reached.}
    \label{fig:precision_percent}
\end{figure}

\subsection{Results: Self Inversion}

Analysis of the precision attained across various network architectures yields several key insights:

\begin{enumerate}
\item An early stopping criterion was placed for trials 0-4, resulting in poor performance in terms of precision (see Table \ref{tab:inversion_table}).   

\item In trial 5, we relaxed the condition of early stopping and trained the network for 1000 epochs. The network further optimizes the weights of the network, reaching a lower value of the objective function, and thus, the inversion precision improves for $\teff$, $\logg$ and $\feh$. 

\item In trial 6, we reduced the number of parameters and increased the training epochs. The training takes a significant amount of time ($\sim$ 28 hours). We reached the lowest values of the objective function with this architecture. The inverted parameters have the highest precision for this case but this ANN model might not be able to generalise the new dataset as it seems to do overfitting. This trial showed inaccurate values of derived physical parameters when the observed spectrum of stars in NGC 6397 was inverted using this network. 

\item Fig. \ref{fig:loss_vs_epoch} illustrates the training curves for different network architectures. Notably, simpler architectures with only two hidden layers (trial 7) exhibit faster training progress (lower training time), as reported in Table \ref{tab:train_table} as compared to trial 6, which has complex architecture. 

\item A consistent trend emerges wherein lower minimum mean residual error (min MSE) correlates with decreased precision (better accuracy) in inverted parameters, regardless of the network architecture. Fig. \ref{fig:precision_percent} visually depicts this relationship. 

% \item Fig. \ref{fig:precision_percent} suggests that training should ideally be stopped when the mean residual variance reaches approximately ($\leq 10^{-6}$) to achieve the precision level of $\equiv 5\%$.

\end{enumerate}

The optimal choice of interpolator based on the above discussion is the interpolator trained in trial 7. Trial 7 is the simplest architecture that is trained for $10^{5}$ epochs, which reached the lowest value of objective in $\sim 7 $ hours. While the total training time is considerable, once the network is trained, the interpolator can generate stellar spectra in real-time by simply inputting the physical parameters. This represents a significant computational advantage over traditional methods, which involve solving model atmosphere codes and can take several days to generate a single spectrum. However, it should be noted that the model is trained on a pre-computed grid of stellar spectra and can be used as an alternative to traditional interpolation techniques. Ideally to get the real theoretical spectrum one would be required to solve the model atmosphere.

%-------------------------------------------------------------------

\section{Determination physical parameters of NGC 6397}\label{sec:ngc6397_physical_parameters}

\begin{figure*}
    \centering
    \includegraphics{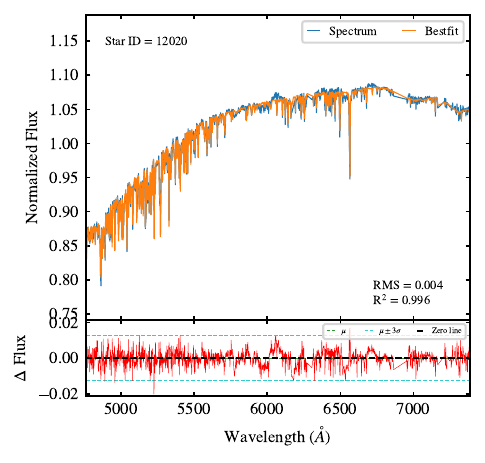}
    \includegraphics{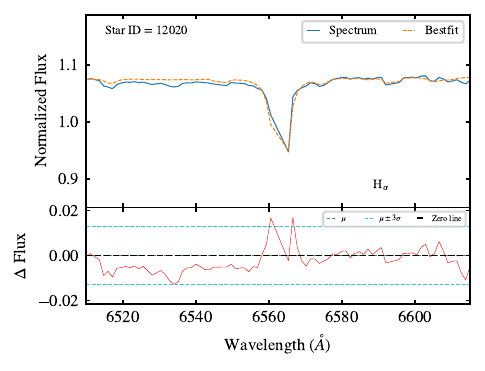}
    \includegraphics{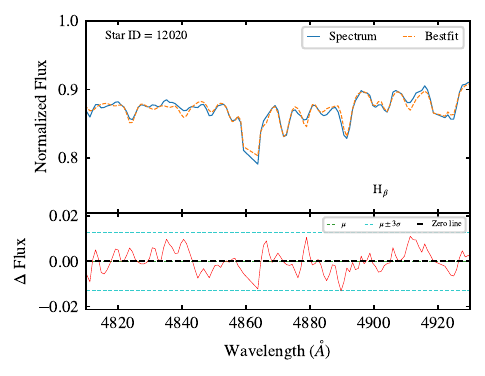}
    \includegraphics{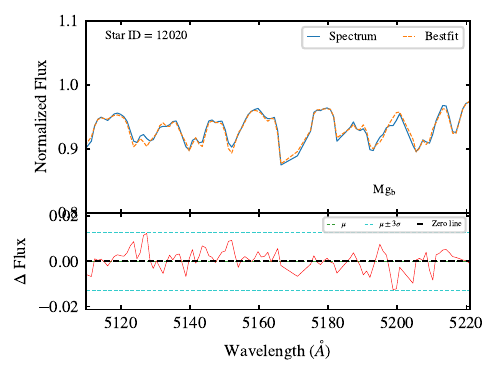}
    \caption{A typical spectrum fits of (ID: id000012020jd2456865p6115f000) is shown with the residuals. The wavelength range of the fit is 4762.25 \AA{} to 7384.75 \AA{}. The physical parameters of this spectrum are determined as $\teff = 5034 \pm 4$ K, $\logg = 1.664 $ (fixed photometric) dex and $\feh = -1.99 \pm 0.01$ dex. The global rms residual for the fit is $0.004$. In the top right and in the bottom panels, we show the fits for regions around $H_\alpha$, H$_\beta$ and Mg$_{\rm b}$ lines. }
    \label{fig:comaprison_spectra}
\end{figure*}

The trained neural network is essentially a function `$f$', which generates the spectrum corresponding to the input atmospheric parameters (\textbf{x}) in a given wavelength range. 
\begin{equation}
    \text{S} = f(\textbf{x}; w),    
\end{equation}

where S is the generated spectrum and $\textbf{x} \equiv [\teff, \logg, \feh]$. $w$ are the weights of the network that were fixed after the training of the network. The atmospheric parameters of stars in NGC 6397 were derived by minimizing the squared residual between the observed and the interpolated spectrum ($F$), 
\begin{equation}
    F = P_n \times G(v_{\rm rad}, \sigma_{\rm rel}) \otimes f(\textbf{x}; w).
\end{equation}

$P_n$ is a series of Legendre polynomials of degree $n$, used to absorb the instrumental response and extinction in line-of-sight. We used $n=20$. The Gaussian function $G(v_{\rm rad}, \sigma_{\rm rel})$ is centered around $v_{\rm rad}$ with the standard deviation $\sigma_{\rm rel}$. For more details, see \cite{arentsen_stellar_2019}. 

A $\chi^2$ minimization is performed between the interpolated spectra and the observed spectra using the Levenberg-Marquardt method in the spectroscopic software ULySS to get the best-fit parameters. As the non-linear least square fitting methods are sensitive to the initial guesses of the parameters, it is essential to provide good prior estimates of parameters. We used three initial guesses for $\teff$, 5000 K, 5500 K and 6000 K. For $\feh$, we used -1.0 dex and -1.5 dex as initial guesses. The final parameters derived by ULySS are provided in Table \ref{tab:final_parameters}.

\begin{figure}
    \centering
    \includegraphics{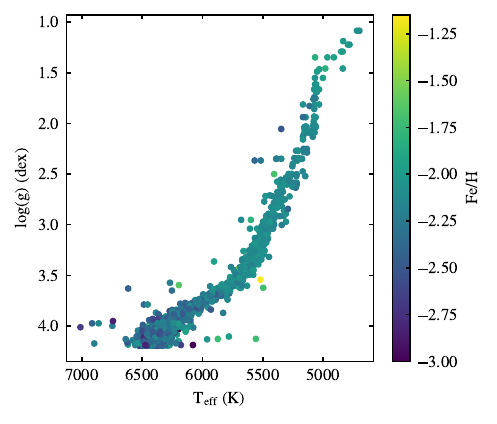}
    \caption{Distribution of derived stellar parameters of stars in NGC 6397 using ANN-spectral interpolator based on medium resolution GSL grid with $\afe$ = +0.4 integrated with ULySS.}
    \label{fig:teff-logg-feh}
\end{figure}

In Fig. \ref{fig:comaprison_spectra}, we show the typical fit for one spectrum (ID: id000012020jd2456865p6115f000). The wavelength range of the fit is from 4762.25 \AA{} to 7384.75 \AA{}. For this particular spectrum, we achieved a global residual rms of 0.004. The derived physical parameters for this particular spectrum are $\teff =5034 \pm 4$ K, $\logg = 1.664$ (fixed photometric) dex and $\feh = -1.99 \pm 0.01$ dex. We have also shown the fits of spectra around regions of $H_\alpha$, H$_\beta$ and Mg$_{\rm b}$ line and the corresponding residual is also shown in the plots. We observe that the ANN interpolator is able to accurately model the spectrum for the star. We have provided the derived atmospheric parameters for 1587 spectra in Table \ref{tab:final_parameters} along with the systematic error provided by the fitting routine. We have also given the RA and Dec information of the stars directly from the spectra fits file and also provided the associated quality flag of each spectrum. Fig. \ref{fig:teff-logg-feh} depicts the derived physical parameters of the stars in the NGC 6397 cluster.

\begin{table*}
    \centering
    \caption{The derived physical parameters are provided for the MUSE spectra of stars in NGC 6397. The parameters are derived using ULySS software, which uses the ANN interpolator trained on the GSL grid of synthetic spectra with $\afe = +0.4$. The ID, RA, Dec and Quality Flag (QF) is adapted from the H16.}
    \label{tab:final_parameters}
    \begin{tabular}{crrcccr}
\hline\hline
      ID &   RA (deg) &  Dec (deg) &  T$_{\rm eff}$ (K) &     log g (dex) &     [Fe/H] (dex) &  QF \\
\hline
id000012020jd2456865p6115f000 & 265.146786 & -53.661912 & 5034 $\pm$ 4 & 1.664 & -1.99 $\pm$ 0.01 & 0 \\
id000009159jd2456865p5898f000 & 265.199043 & -53.662589 & 5002 $\pm$ 7 & 1.551 & -1.94 $\pm$ 0.01 & 0 \\
id000004871jd2456865p6043f000 & 265.188475 & -53.685418 & 5136 $\pm$ 9 & 2.251 & -2.24 $\pm$ 0.01 & 0 \\
id000009050jd2456865p5898f000 & 265.200481 & -53.671457 & 4720 $\pm$ 15 & 1.085 & -2.07 $\pm$ 0.02 & 0 \\
id000006692jd2456865p5970f000 & 265.162076 & -53.682601 & 4796 $\pm$ 15 & 1.222 & -2.01 $\pm$ 0.02 & 0 \\
\vdots & \vdots  & \vdots  &  \vdots   & \vdots  & \vdots  &   \vdots \\
id000014699jd2456866p5251f000 & 265.161285 & -53.650406 & 6508 $\pm$ 48 & 4.154 & -3.0 $\pm$ -0.0 & 0 \\
id000015700jd2456870p5705f000 & 265.189309 & -53.640135 & 6141 $\pm$ 73 & 4.056 & -1.95 $\pm$ 0.15 & 0 \\
id000005796jd2456865p5826f000 & 265.175496 & -53.678024 & 6428 $\pm$ 61 & 4.150 & -2.47 $\pm$ 0.27 & 0 \\
id000005692jd2456865p5826f000 & 265.176501 & -53.674819 & 6326 $\pm$ 68 & 4.171 & -2.85 $\pm$ 0.31 & 0 \\
id000005811jd2456866p4985f000 & 265.175659 & -53.674599 & 5558 $\pm$ 115 & 4.127 & -1.73 $\pm$ 0.17 & 0 \\
\hline
\end{tabular}\\
% \tablefoot{The $\logg$ values are fixed to the photometric values and are adopted from H16.}
{\hspace{-5cm}Note: The $\logg$ values are fixed to the photometric values and are adopted from H16.}
\end{table*}

\begin{figure*}
    \centering
    \includegraphics{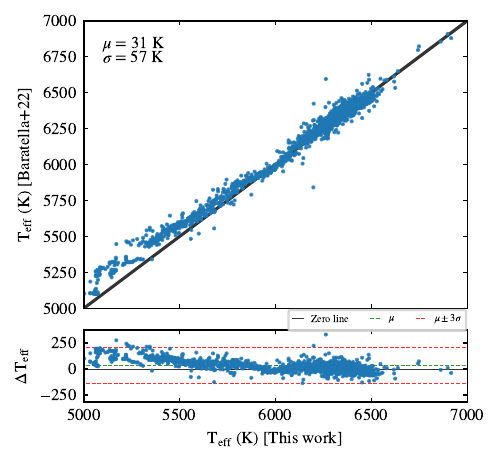}
    \includegraphics{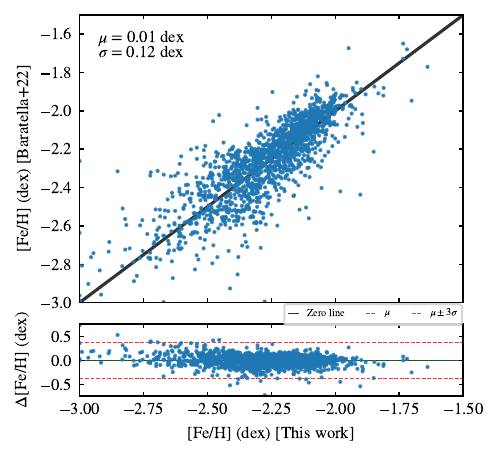}
    \caption{The comparison of derived $\teff$ and $\feh$ with the parameters derived by \citealt{baratella_prospects_2022} using GSL grid with cubic spline interpolation. The black solid line in each plot represents the unit slope line. }
    \label{fig:atm-param-comparison}
\end{figure*}

In Figure \ref{fig:atm-param-comparison}, we compare the parameters ($\teff$ and $\feh$) derived using ANN spectral interpolator (trial 7) of GSL grid ($\afe$ = +0.4) within ULySS with those of B22 using the same grid of synthetic models with same $\afe$ but with cubic Bezi\'er splines interpolation scheme using the analysis software FERRE. Since B22 fixed the surface gravity to their photometric values adopted from H16, we also fixed the $\logg$ values for the sake of comparison. For $\teff$, we observe a mean difference ($< \teff (B22) -  \teff (TW) >$) of 31 K with a standard deviation of 57 K between the two values. The comparison of $\feh$ yields the average difference in ${\feh}$ $ \equiv $ 0.01 dex and the standard deviation of the difference in ${\feh} \equiv 0.12$ dex. The derived values of $\teff$ are in excellent agreement with the values of B22 except at the lower temperature range than the values for $\teff \lessapprox 5700 K$, where we observe a systematic offset. This discrepancy may arise from non-local thermodynamic equilibrium (NLTE) effects, which become significant for metal-poor stars at lower temperatures \citep{Klevas2016A&A...586A.156K}. The $\feh$ shows a significant spread in the derived values.

%-------------------------------------------------------------------

\section{Discussion}\label{sec:discussion}

A number of interpolation schemes exist and are being used in different analysis software. For example, in FERRE, there are three distinct interpolation schemes, namely linear, quadratic and cubic Bezi\'er splines \citep{auer_formal_2003}. This mechanism performs local interpolations using two nearest neighbours for linear interpolation and four nearest neighbours for cubic interpolation. The accuracy of different interpolation schemes was assessed by \cite{meszaros_interpolation_2013} by computing spectra at random locations between the parameter space and comparing them with the interpolated spectra. The cubic interpolation results in the residuals that are about 40\% smaller than the linear interpolation. ULySS uses a totally different approach for the interpolation of the given grid. A polynomial-based spectral interpolator as a function of $\teff$, $\logg$ and $\feh$ is fitted based on the physical parameters for empirical libraries such as MILES and ELODIE \citep{koleva_ulyss_2009, prugniel_atmospheric_2011, sharma_new_2016}.

\subsection{Mean Metallicity}
NGC 6397 has the metallicity derived as $\feh$ = -2.02 dex by \cite{harris_new_2010}. The high-resolution (R $\sim$ 43000) spectra of 5 TO stars and 3 subgiants were analysed by \cite{gratton_o-na_2001} in the metallicity range -2.10 $< \feh <$ -2.00 dex with a precision of about 0.04 dex.

In our analysis, we measured the mean metallicity of stars in NGC 6397 as $< \feh >$ = -2.26 $\pm$ 0.19. This value of $\feh$ is derived for the fixed $\afe$ = +0.4. B22 derived the range of mean metallicity as -2.034 to -2.3 with the GSL grid and with the same fixed $\afe$ = +0.4 with the cubic Bezi\'er interpolation scheme in the same wavelength range. \cite{jain_ngc_2020} used the MILES and ELODIE spectral libraries to derive the metallicities of stars in the cluster using the same spectra and wavelength range. The $\afe$ of these empirical spectral libraries is similar to the solar neighbourhood as most of the stars in these libraries are from the solar neighbourhood. The mean $\feh$ is derived to be $-2.09$ dex with a spread of about 0.08 dex.    

% \noten{REWRITE}
% As we have used the same data for the derivation of parameters except a difference in the analysis of the data. We have used the same synthetic grid and same $\afe$ values except the interpolation scheme. In the analysis of B22 the interpolation method did not have any significant impact on the mean metallicity of the cluster. However we report a difference of about $0.16$ dex between our results and B22.

\subsection{Revisiting Metallicity Trend}

\begin{figure*}
    \centering
    \includegraphics{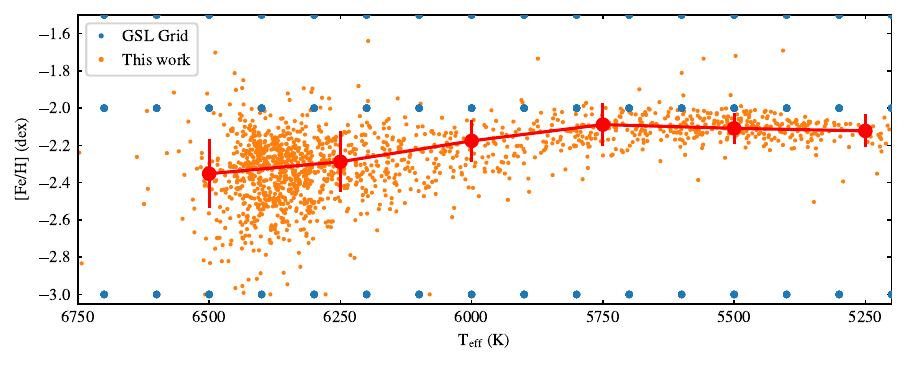}
    \caption{The derived metallicity is plotted against the effective temperature for giants ($\logg <$ 4.2 dex). The solid red line connects the mean and standard deviations of the metallicity in 250 K wide bins. The blue dots represent the node points of the GSL grid.}
    \label{fig:metallicity-trend}
\end{figure*}

H16 found a metallicity trend along the isochrone with an amplitude of 0.2 to 0.3 dex. They found that the metallicity is lowest at the tun-off (TO) and rises both along the main sequence (MS) and sub-giant branch (sGB). A 0.1 dex variation in metallicity between TO and sGB was reported by \cite{korn_atomic_2007} and by \cite{nordlander_atomic_2012} from their theoretical models for NGC 6397. They suggest that atomic fusion plays an important role in the case of metal-poor stars. For a star in a main sequence with low metallicity, heavy elements sink to deeper layers of the atmosphere and disappear from the spectrum. When the star reaches the turn-off point, the convection starts and transports the heavy elements back up at the surface, and thus, the surface metallicity should increase. \cite{jain_ngc_2020} re-analyzed the spectra of the same stars with two different empirical spectral libraries (MILES \& ELODIE) and found no trend of metallicity with ELODIE and a trend opposite to that of H16 with MILES. They also discussed the importance of interpolation schemes when deriving the iron abundances. 

In Figure \ref{fig:metallicity-trend}, we show the metallicity trend along the TO to sGB. We binned the values of metallicity along the temperature with a bin size of 250 K. In our analysis, we report $\feh$ at TO be -2.35 $\pm$ 0.19 dex and at the sGB to be -2.12 $\pm$ 0.09 dex. Hence, we see a metallicity variation of about 0.23 dex, which coincides with the upper limit of variation reported by H16. 

\section{Conclusion}\label{sec:conclusion}

We trained multiple ANN architectures to interpolate the stellar spectra within a given grid of atmospheric parameters. These neural networks used a standardized scaling method for input parameters, ensuring a mean of zero and unit standard deviation across all dimensions in parameter space. This preprocessing step is crucial for maintaining the stability of the network and providing reliable outputs. We utilized the Adam optimization algorithm, which iteratively updated the network parameters to minimize the mean squared error (MSE). Notably, our analysis revealed that the complexity of the network and the training duration had a substantial impact on the precision of the interpolated parameters.

Our study involved nine distinct training trials, each varying in network architecture and training iterations. Early stopping was employed in the initial trials, halting training when the loss function plateaued, which resulted in suboptimal precision. For instance, trials 0-4, which employed early stopping, showed poor performance in terms of precision. Relaxing this criterion in trial 5 and extending training to 1000 epochs allowed for further optimization of network weights, leading to improved precision for effective temperature ($\teff$), surface gravity ($\logg$), and metallicity ($\feh$). This trial, however, still highlighted the balance between model complexity and overfitting, as seen in trial 6, where increased training epochs and reduced parameters led to overfitting despite achieving the lowest MSE.

We integrated the ANN interpolator with the University of Lyon Spectroscopic Analysis Software (ULySS) to facilitate parameter retrieval from observed spectra. The precision of the interpolator was quantified by the residual variance between the actual and derived parameters, with our results indicating that the precision was consistently better with lower MSE values. This relationship between MSE and precision is a critical insight for future applications of neural networks in spectroscopic analysis.

The choice of interpolator is important, as shown by our findings that the simplest architecture (trial 7), trained for 100,000 epochs, provided the most practical balance of training time and precision. This model, which required around seven hours of training time, significantly reduces computational overhead compared to traditional methods, which often take days to generate single spectra. 

Applying our ANN-based methodology to NGC 6397, we derived atmospheric parameters for 1587 spectra and compared them with those from previous studies. The derived effective temperatures ($\teff$) showed excellent concordance with values from B22, with a mean difference of just 31 K and a standard deviation of 57 K. The metallicity ($\feh$) values displayed a mean difference of 0.01 dex and a standard deviation of 0.12 dex. This variation highlights the sensitivity of metallicity measurements to different interpolation schemes and underscores the importance of methodological consistency in stellar studies.

In terms of mean metallicity, our findings for NGC 6397 revealed a mean $\feh$ of -2.26 $\pm$ 0.19, slightly lower than previous estimates of -2.02 dex by \cite{harris_new_2010}. This discrepancy, while within the range reported by B22 (-2.034 to -2.3 dex), highlights the need for careful consideration of interpolation methods and grid completeness in spectral analysis. The broad spread of our metallicity values underscores the complexity of accurately measuring this parameter in metal-poor environments.

Revisiting the metallicity trend along the isochrone, we observed a significant variation of approximately 0.23 dex between the turn-off (TO) and sub-giant branch (sGB) stars. This trend, which aligns with the upper limit of variation reported by H16, emphasizes the dynamic nature of metallicity in stellar atmospheres. Our analysis indicates that the metallicity at the TO is about -2.35 $\pm$ 0.19 dex, increasing to -2.12 $\pm$ 0.09 at the sGB.

The ability to generate stellar spectra in real time from physical parameters offers a significant computational advantage over traditional methods. However, the generated stellar spectra would be as \textit{real} as the training grid. The insights gained from our analysis also highlight the importance of methodological consistency and the careful selection of interpolation schemes in deriving accurate stellar parameters. As we continue to refine these techniques, the integration of ANN-based spectral interpolators with spectroscopic analysis software like ULySS promises to enhance our understanding of stellar populations and their underlying physical properties.

\section*{Acknowledgements}
The authors acknowledge the usage of the IUCAA HPC computing facility for numerical calculations (training ANN models). NK acknowledges the financial assistance from the Council of Scientific and Industrial Research (CSIR), New Delhi, India, as the Senior Research Fellowship (SRF) file no. 09/45(1651)/2019-EMR-I.

\bibliographystyle{elsarticle-harv} 
\bibliography{mybib}



%% For example:

%% \bibitem[Aladro et al.(2015)]{Aladro15} Aladro, R., Martín, S., Riquelme, D., et al. 2015, \aas, 579, A101

%%\end{thebibliography}

\end{document}